\documentclass[preprint,aps, showpacs]{revtex4}
\usepackage{graphicx}
\begin{document}
\draft
\title{Partial Wave Analysis of Scattering with Nonlocal Aharonov-Bohm Effect and
Anomalous Cross Section Induced by Quantum Interference}
\author{De-Hone Lin\thanks{%
e-mail: dhlin@mail.nsysu.edu.tw,}}
\address{Department of Physics\\
National Sun Yat-sen University, Kaohsiung, Taiwan}
\date{\today }
\begin{abstract}
Partial wave theory of a three dmensional scattering problem for an arbitray
short range potential and a nonlocal Aharonov-Bohm magnetic flux is
established. The scattering process of a ``hard shere'' like potential and
the magnetic flux is examined. An anomalous total cross section is revealed
at the specific quantized magnetic flux at low energy which helps explain
the composite fermion and boson model in the fractional quantum Hall effect.
Since the nonlocal quantum interference of magnetic flux on the charged
particles is universal, the nonlocal effect is expected to appear in quite
general potential system and will be useful in understanding some other
phenomena in mesoscopic phyiscs.
\end{abstract}
\maketitle
\tolerance=10000
\section{Introduction}

Since the global structure of magnetic flux was discovered about 40 years
ago \cite{1}, it had great contribution to our comprehension of the
foundation of quantum theory \cite{2}, the phenomenon of quantum Hall effect 
\cite{3}, superconductivity \cite{4}, repulsive Bose gases \cite{6}, and,
recently, help to explore the quantum computers, quantum cryptography
communication systems \cite{7,8}. Nevertheless, to my knowledge, a general
partial wave analysis for a scattering of a charged particle moving in an
arbitrary short range potential plus a magnetic flux in three dimensions is
still not done until now \cite{8a}. In this paper we discuss partial wave
method of a charged particle moving in an arbitrary short range potential
with scattering center located at the origin, and the AB magnetic flux along
z-axis in the three dimensional space. Special attention is paid to the
problem of the ``hard sphere'' like potential plus the magnetic flux with
the incident direction of particles restricted in $x$-$y$ plane. Several
interesting results are concluded as follows: (1) In the long wave length
limit ( equivalently, short range potential) the total cross section is
drastically suppressed at quantized magnetic flux $\Phi =(2n+1)\Phi _{0}/2,$
where $n=0,1,2,\cdots $, and $\Phi _{0}$ is the fundamental magnetic flux
quantum $hc/e$. The global influence of the magnetic flux on the cross
section is manifested with $\Phi _{0}$ periodicity. The result provides
another possibility to explain the anomalous total cross section given in
Ref. \cite{9}, where the quantum entanglement is supposedly responsible for
the suppression of total cross section in the condensed system. On the other
hand, the cross section approaches the flux-free case in the short wave
length limit, i.e. the quantum interference feature of the nonlocal effect
gradually disappears, and the cross section approaches the classical limit.
(2) If the hard sphere is used to simulate the boson (fermion) moving in $x$-%
$y$ plane, the scattering process of identical particles carrying the
magnetic flux shows that the total cross section is suppressed at quantized
magnetic flux $\Phi =(2n+1)\Phi _{0}$ for bosons ($\Phi =2n\Phi _{0}$ for
fermions) and exhibits the global structure with $2\Phi _{0}$ periodicity.
These results shed light on the model of composite bosons and fermions in
the fractional quantum Hall effect \cite{3,11,12}, superconductivity, and
transport phenomena in nanostructures \cite{13a,13b}. Furthermore, since the
nonlocal influence of the magnetic flux on the charged particle are
universal, the implication should be general in similar systems.

This paper is organized as follows. In section II, the partial wave method
of scattering with AB effect in three dimensions is established. The
nonintegrable phase factor (NPF) \cite{13} is used to couple the magnetic
flux with the particle angular momentum such that the partial wave method
can be conveniently developed. In section III, special attention is paid to
the specific condition of the incident direction restricted in the $x$-$y$
plane. The total cross section of a charged particle with its path in $x$-$y$
plane scattered by a hard sphere potential plus an AB magnetic flux is
discussed in some detail. Our discussions are summarized in section V.

\section{Partial Wave Analysis of Scattering With Nonlocal Aharonov-Bohm
Effect}

We consider a three-dimensional model. The fixed-energy Green's function $%
G^{(0)}({\bf x,x}^{\prime };E)$ for a charged particle with mass $\mu $
propagating from ${\bf x}^{\prime }$ to ${\bf x}$ satisfies the
Schr\"{o}dinger equation 
\begin{equation}
\left\{ E-\left[ -\frac{\hbar ^{2}\nabla ^{2}}{2\mu }+V({\bf x})\right]
\right\} G^{(0)}({\bf x,x}^{\prime };E)=\delta ({\bf x-x}^{\prime }),
\label{01}
\end{equation}
where $V({\bf x})$ is the scalar potential and ${\bf x}$ is the three
dimensional coordinate vector. In the spherically symmetric system, the
Green's function can be decomposed as \cite{13c} 
\begin{equation}
G^{(0)}({\bf r,r}^{\prime };E)=\sum_{l=0}^{\infty
}\sum_{m=-l}^{l}G_{l}^{(0)}(r{\bf ,}r^{\prime };E)Y_{lm}(\theta ,\varphi
)Y_{lm}^{\ast }(\theta ^{\prime },\varphi ^{\prime })  \label{02}
\end{equation}
with $Y_{lm}(\theta ,\varphi )$ the well-known spherical harmonics and $%
G_{l}^{(0)}(r{\bf ,}r^{\prime };E)$ the radial Green's function for the
specific angular momentum channel $l$. The left-hand side of Eq. (\ref{01})
can then be cast into 
\[
\sum_{l=0}^{\infty }\sum_{m=-l}^{l}\left\{ E+\left[ \frac{\hbar ^{2}}{2\mu }%
\left( \frac{d^{2}}{dr^{2}}+\frac{2}{r}\frac{d}{dr}\right) -\frac{%
l(l+1)\hbar ^{2}}{2\mu r^{2}}\right] -V(r)\right\} 
\]
\begin{equation}
\times G_{l}^{(0)}(r{\bf ,}r^{\prime };E)Y_{lm}(\theta ,\varphi
)Y_{lm}^{\ast }(\theta ^{\prime },\varphi ^{\prime }).  \label{03}
\end{equation}
For a charged particle affected by a magnetic field, the Green's function $G(%
{\bf x,x}^{\prime };E)$ is different from $G^{(0)}({\bf x,x}^{\prime };E)$
by a global NPF \cite{13,13c,14,15,16,17,18} 
\begin{equation}
G({\bf x},{\bf x}^{\prime };E)=G^{(0)}({\bf x},{\bf x}^{\prime };E)\exp
\left\{ \frac{ie}{\hbar c}\int_{{\bf x}^{\prime }}^{{\bf x}}{\bf A(\tilde{x}}%
)\cdot d{\bf \tilde{x}}\right\} .  \label{04}
\end{equation}
Here the vector potential ${\bf A(x})$ is used to represent the magnetic
field. For an infinitely thin tube of finite magnetic flux along the $z$%
-direction, the vector potential can be described by 
\begin{equation}
{\bf A(x})=2g\frac{-y{\hat{e}}_{x}+x{\hat{e}}_{y}}{x^{2}+y^{2}},  \label{05}
\end{equation}
where ${\hat{e}}_{x},{\hat{e}}_{y}$ stand for the unit vector along the $x,y$
axis respectively. Introducing the azimuthal angle $\varphi ({\bf x})=\tan
^{-1}(y/x)$ around the AB tube, the components of the vector potential can
be expressed as $A_{i}=2g\partial _{i}\varphi ({\bf x}).$ The associated
magnetic field lines are confined to an infinitely thin tube along the $z$%
-axis, 
\begin{equation}
B_{3}=2g\epsilon _{3ij}\partial _{i}\partial _{j}\varphi ({\bf x})=4\pi
g\delta ({\bf x}_{\bot }),  \label{06}
\end{equation}
where ${\bf x}_{\bot }${\bf \ }stands for the transverse vector ${\bf x}%
_{\bot }\equiv (x,y).$ Since the magnetic flux through the tube is defined
by the integral $\Phi =\int d^{2}xB_{3}$, the coupling constant $g$ is
related to the magnetic flux by $g=\Phi /4\pi $. By using the expression of $%
A_{i}=2g\partial _{i}\varphi $, the angular difference between the initial
point ${\bf x}^{\prime }$ and the final point ${\bf x}$ in the exponent of
the NPF is given by 
\begin{equation}
\varphi -\varphi ^{\prime }=\int_{t}^{t^{\prime }}d\tau \dot{\varphi}(\tau
)=\int_{t}^{t^{\prime }}d\tau \frac{-y\dot{x}+x\dot{y}}{x^{2}+y^{2}}=\int_{%
{\bf x}^{\prime }}^{{\bf x}}\frac{{\bf \tilde{x}\times }d{\bf \tilde{x}}}{%
{\bf \tilde{x}}^{2}},  \label{07}
\end{equation}
where $\dot{\varphi}=d\varphi /d\tau $. Given two paths $C_{1}$ and $C_{2}$
connecting ${\bf x}^{\prime }$ and ${\bf x}$, the integral differs by an
integer multiple of $2\pi $. The winding number is thus given by the contour
integral over the closed difference path $C$: 
\begin{equation}
n=\frac{1}{2\pi }\oint_{C}\frac{{\bf \tilde{x}\times }d{\bf \tilde{x}}}{{\bf 
\tilde{x}}^{2}}.  \label{08}
\end{equation}
The magnetic interaction is therefore purely nonlocal and topological \cite
{18a}. Its action takes the form ${\cal A}_{{\rm mag}}=-\hbar \mu _{0}2\pi
n, $ where $\mu _{0}\equiv -2eg/\hbar c=-\Phi /\Phi _{0}$ is a dimensionless
number with the customarily minus sign. The NPF now becomes $\exp \left\{
-i\mu _{0}(2\pi n+\varphi -\varphi ^{\prime })\right\} $. With the help of
the equality between the associated Legendre polynomial $P_{l}^{m}(z)$ and
the Jacobi function $P_{n}^{\left( \alpha ,\beta \right) }(z)$ \cite{17,18}, 
\begin{equation}
P_{l}^{m}(\cos \theta )=(-1)^{m}\frac{\Gamma (l+m+1)}{\Gamma (m+1)}\left(
\cos \frac{\theta }{2}\sin \frac{\theta }{2}\right) ^{m}P_{l-m}^{\left(
m,m\right) }(\cos \theta ),  \label{09}
\end{equation}
the angular part of the Green's function in the expression (\ref{02}) can be
turned into the following form 
\[
\sum_{m{\bf =-}l}^{l}Y_{lm}(\theta ,\varphi )Y_{lm}^{\ast }(\theta ^{\prime
},\varphi ^{\prime })=\sum_{m{\bf =-}l}^{l}\frac{2l+1}{4\pi }\frac{\Gamma
\left( l-m+1\right) }{\Gamma \left( l+m+1\right) }P_{l}^{m}(\cos \theta
)P_{l}^{m}(\cos \theta ^{\prime })e^{im(\varphi -\varphi ^{\prime })} 
\]
\[
=\sum_{m{\bf =-}l}^{l}\left[ \frac{2l+1}{4\pi }\frac{\Gamma \left(
l-m+1\right) \Gamma \left( l+m+1\right) }{\Gamma ^{2}\left( l+1\right) }%
\right] \left( \cos \frac{\theta }{2}\cos \frac{\theta ^{\prime }}{2}\sin 
\frac{\theta }{2}\sin \frac{\theta ^{\prime }}{2}\right) ^{m} 
\]
\begin{equation}
\times P_{l-m}^{\left( m,m\right) }(\cos \theta )P_{l-m}^{\left( m,m\right)
}(\cos \theta ^{\prime })e^{im\left( \varphi -\varphi ^{\prime }\right) }.
\label{010}
\end{equation}
In order to include the NPF due to the AB effect, we will change the index $%
l $ into $q$ related by the definition $l-m=q$. As a result Eq. (\ref{03})
can be rewritten as 
\[
\sum_{q=0}^{\infty }\sum_{m=-\infty }^{\infty }\left\{ E+\left[ \frac{\hbar
^{2}}{2\mu }\left( \frac{d^{2}}{dr^{2}}+\frac{2}{r}\frac{d}{dr}\right) -%
\frac{(q+m)(q+m+1)\hbar ^{2}}{2\mu r^{2}}\right] -V(r)\right\} 
\]
\[
\times G_{q+m}^{(0)}(r{\bf ,}r^{\prime };E)\left[ \frac{2(q+m)+1}{4\pi }%
\frac{\Gamma \left( q+1\right) \Gamma \left( q+2m+1\right) }{\Gamma
^{2}\left( q+m+1\right) }\right] \left( \cos \frac{\theta }{2}\cos \frac{%
\theta ^{\prime }}{2}\sin \frac{\theta }{2}\sin \frac{\theta ^{\prime }}{2}%
\right) ^{m} 
\]
\begin{equation}
\times P_{q}^{\left( m,m\right) }(\cos \theta )P_{q}^{\left( m,m\right)
}(\cos \theta ^{\prime })e^{im\left( \varphi -\varphi ^{\prime }\right) }.
\label{011}
\end{equation}
The Green's function $G_{n}(r,r^{\prime };E)$ for a specific winding number $%
n$ can be obtained by converting the summation over $m$ in Eq. (\ref{011})
into an integral over $z$ and another summation over $n$ by the Poisson's
summation formula (e.g. Ref. \cite{19} p.469) 
\begin{equation}
\sum_{m=-\infty }^{\infty }f(m)=\int_{-\infty }^{\infty }dz\sum_{n=-\infty
}^{\infty }e^{2\pi nzi}f(z).  \label{012}
\end{equation}
So the expression (\ref{03}) when includes the NPF can be written as 
\[
\sum_{q=0}^{\infty }\int dz\sum_{n=-\infty }^{\infty }\left\{ E+\left[ \frac{%
\hbar ^{2}}{2\mu }\left( \frac{d^{2}}{dr^{2}}+\frac{2}{r}\frac{d}{dr}\right)
+\frac{(q+z)(q+z+1)\hbar ^{2}}{2\mu r^{2}}\right] -V(r)\right\} 
\]
\[
\times G_{q+z}(r{\bf ,}r^{\prime };E)\left[ \frac{2(q+z)+1}{4\pi }\frac{%
\Gamma \left( q+1\right) \Gamma \left( q+2z+1\right) }{\Gamma ^{2}\left(
q+z+1\right) }\right] \left( \cos \frac{\theta }{2}\cos \frac{\theta
^{\prime }}{2}\sin \frac{\theta }{2}\sin \frac{\theta ^{\prime }}{2}\right)
^{z} 
\]
\begin{equation}
\times P_{q}^{\left( z,z\right) }(\cos \theta )P_{q}^{\left( z,z\right)
}(\cos \theta ^{\prime })e^{i(z-\mu _{0})\left( \varphi +2n\pi -\varphi
^{\prime }\right) },  \label{013}
\end{equation}
where the superscript $(0)$ in $G_{q+m}^{(0)}$ has been suppressed to denote
that the AB effect is included. Obviously, the number $n$ in the right-hand
side is precisely the winding number by which we want to classify the
Green's function. Employing the special case of the Poisson formula $%
\sum_{n=-\infty }^{\infty }\exp \{ik(\varphi +2n\pi -\varphi ^{\prime
})\}=\sum_{m=-\infty }^{\infty }\delta (k-m)\exp \{im(\varphi -\varphi
^{\prime })\},$ the summation over all indices $n$ forces $z=\mu _{0}$
modulo an arbitrary integer number. Thus, we obtain 
\[
\sum_{q=0}^{\infty }\sum_{m=-\infty }^{\infty }\left\{ E+\left[ \frac{\hbar
^{2}}{2\mu }\left( \frac{d^{2}}{dr^{2}}+\frac{2}{r}\frac{d}{dr}\right) -%
\frac{(q+\left| m+\mu _{0}\right| )(q+\left| m+\mu _{0}\right| +1)\hbar ^{2}%
}{2\mu r^{2}}\right] -V(r)\right\} 
\]
\[
\times G_{q+\left| m+\mu _{0}\right| }(r{\bf ,}r^{\prime };E)\left\{ \frac{%
\left[ 2\left( q+\left| m+\mu _{0}\right| \right) +1\right] }{4\pi }\frac{%
\Gamma \left( q+1\right) \Gamma \left( 2\left| m+\mu _{0}\right| +q+1\right) 
}{\Gamma ^{2}\left( \left| m+\mu _{0}\right| +q+1\right) }\right\}
e^{im\left( \varphi -\varphi ^{\prime }\right) } 
\]
\begin{equation}
\times \left( \cos \frac{\theta }{2}\cos \frac{\theta ^{\prime }}{2}\sin 
\frac{\theta }{2}\sin \frac{\theta ^{\prime }}{2}\right) ^{\left| m+\mu
_{0}\right| }P_{q}^{\left( \left| m+\mu _{0}\right| ,\left| m+\mu
_{0}\right| \right) }(\cos \theta )P_{q}^{\left( \left| m+\mu _{0}\right|
,\left| m+\mu _{0}\right| \right) }(\cos \theta ^{\prime }).  \label{014}
\end{equation}
We see that the influence of the AB effect to the radial Green's function is
to replace the integer quantum number $l$ with a real one $(q+\left| m+\mu
_{0}\right| )$ which depends on the magnitude of magnetic flux. Analogously
the same procedure can be applied to the delta function $\delta ({\bf r-r}%
^{\prime })$ in the \ r.h.s. of Eq. (\ref{01}) by employing the solid angle
representation of the $\delta $ function 
\begin{equation}
\delta \left( \Omega -\Omega ^{\prime }\right) =\sum_{l=0}^{\infty
}\sum_{m=-l}^{l}Y_{lm}(\theta ,\varphi )Y_{lm}^{\ast }(\theta ^{\prime
},\varphi ^{\prime }).  \label{015}
\end{equation}
With the help of orthogonal property of the angular part \cite{18}, 
\[
\int_{0}^{2\pi }d\varphi \int_{-1}^{1}(d\cos \theta )P_{q}^{\left( \left|
m+\mu _{0}\right| ,\left| m+\mu _{0}\right| \right) }(\cos \theta
)P_{q^{\prime }}^{\left( \left| m^{\prime }+\mu _{0}\right| ,\left|
m^{\prime }+\mu _{0}\right| \right) }(\cos \theta ) 
\]
\[
\left( \cos \frac{\theta }{2}\sin \frac{\theta }{2}\right) ^{\left| m+\mu
_{0}\right| }\left( \cos \frac{\theta }{2}\sin \frac{\theta }{2}\right)
^{\left| m^{\prime }+\mu _{0}\right| }e^{i(m-m^{\prime })\varphi } 
\]
\begin{equation}
=\frac{\Gamma ^{2}\left( q+\left| m+\mu _{0}\right| +1\right) }{\Gamma
\left( q+1\right) \Gamma \left( q+2\left| m+\mu _{0}\right| +1\right) }\frac{%
4\pi }{2\left( q+\left| m+\mu _{0}\right| \right) +1}\delta _{q,q^{\prime
}}\delta _{m,m^{\prime }},  \label{016}
\end{equation}
one can show that the radial Green's function for the set of the fixed
quantum numbers $(q,m)$ satisfies 
\[
\left\{ E+\left[ \frac{\hbar ^{2}}{2\mu }\left( \frac{d^{2}}{dr^{2}}+\frac{2%
}{r}\frac{d}{dr}\right) -\frac{\alpha (\alpha +1)\hbar ^{2}}{2\mu r^{2}}%
\right] -V(r)\right\} 
\]
\begin{equation}
\times G_{\alpha }(r{\bf ,}r^{\prime };E)=\delta (r{\bf -}r^{\prime }).
\label{018}
\end{equation}
Here we have defined $\alpha \equiv (q+\left| m+\mu _{0}\right| )$ for
convenience. The corresponding radial wave equation then reads 
\begin{equation}
\left[ \frac{d^{2}}{dr^{2}}+\frac{2}{r}\frac{d}{dr}+\left( k^{2}-U(r)-\frac{%
\alpha (\alpha +1)}{r^{2}}\right) \right] R_{k\alpha }(r)=0,  \label{019}
\end{equation}
where $U(r)\equiv 2\mu V(r)/\hbar ^{2}$ and the subscript set $(k,\alpha )$
with $k\equiv \sqrt{2\mu E}/\hbar $ in the radial wave function $R_{k\alpha
}(r)$ denotes the state of scattering particle. For a short range potential,
say $V(r)$ vanishes as $r>a$, the exterior solution is the linear
combination of 1st and 2nd kind spherical Bessel functions $j_{\alpha }(kr)$%
, and $n_{\alpha }(kr)$, and may be given by 
\[
R_{\alpha k}(r)=\left[ C_{\alpha }(k)j_{\alpha }(kr)+D_{\alpha }(k)n_{\alpha
}(kr)\right] 
\]
\begin{equation}
=A_{\alpha }(k)\left[ \cos \delta _{\alpha }(k)j_{\alpha }(kr)-\sin \delta
_{\alpha }(k)n_{\alpha }(kr)\right] ,  \label{020}
\end{equation}
where $\delta _{\alpha }(k)$ is the phase shift defined by $-D_{\alpha
}(k)/C_{\alpha }(k)\equiv \tan \delta _{\alpha }(k)$ and $A_{\alpha }(k)=$ $%
C_{\alpha }(k)/\cos \delta _{\alpha }(k)$ which can be used to measure the
interaction strength of potential. Thus the general solution $\Psi _{{\bf k}%
}({\bf x})$ of a scattering particle with arbitrary incident direction $%
(\theta ^{\prime },\varphi ^{\prime })$ is given by superposition of partial
waves $\Psi _{k\alpha }(r)$, which reads 
\begin{equation}
\Psi _{{\bf k}}({\bf x})=\sum_{q=0}^{\infty }\sum_{m=-\infty }^{\infty
}A_{\alpha }(k)\left[ \cos \delta _{\alpha }(k)j_{\alpha }(kr)-\sin \delta
_{\alpha }(k)n_{\alpha }(kr)\right] {\cal Y}_{qm}^{\ast }(\theta ^{\prime
},\varphi ^{\prime }){\cal Y}_{qm}(\theta ,\varphi )  \label{021}
\end{equation}
in which ${\cal Y}_{qm}(\theta ,\varphi )$ is defined by 
\[
{\cal Y}_{qm}(\theta ,\varphi )=\sqrt{\frac{\Gamma (q+1)\Gamma (\alpha
+\left| m+\mu _{0}\right| +1)}{\Gamma ^{2}(\alpha +1)}} 
\]
\begin{equation}
\times \left( \cos \frac{\theta }{2}\sin \frac{\theta }{2}\right) ^{\left|
m+\mu _{0}\right| }P_{q}^{(\left| m+\mu _{0}\right| ,\left| m+\mu
_{0}\right| )}(\cos \theta )e^{im\varphi }.  \label{022}
\end{equation}
Since it must describe both the incident and the scattered waves at large
distance, we naturally expect it to become 
\begin{equation}
\Psi _{{\bf k}}({\bf x})\stackrel{\left| {\bf x}\right| \rightarrow \infty }{%
\longrightarrow }{\cal F}_{\infty }\left( \exp \{i{\bf k\cdot x}\}\exp
\left\{ \frac{ie}{\hbar c}\int_{C}^{{\bf x}}{\bf A(\tilde{x}})\cdot d{\bf 
\tilde{x}}\right\} \right) +f(\theta ,\varphi )\frac{\exp \{ikr\}}{r},
\label{023}
\end{equation}
where $\exp \{i{\bf k\cdot x}\}$ describes the incident plane wave of a
charged particle with momentum{\bf \ }${\bf p}=\mu {\bf k}$ and ${\cal F}%
_{\infty }(\cdot )$ stands for its asymptotic form. The phase modulation of
the NPF comes from the fact that the field ${\bf A(x})$ of AB magnetic flux
affects the charged particle globally. The subscript $C$ in the integral is
used to represent the nature of the NPF which depends on the different
paths. To find the amplitude $f(\theta ,\varphi )$ we first note that the
plane wave in Eq. (\ref{023}) can be expanded in terms of the spherical
harmonics 
\begin{equation}
e^{i{\bf k\cdot x}}=\sum_{l=0}^{\infty }\sum_{m=-l}^{l}4\pi
i^{l}j_{l}(kr)Y_{lm}^{\ast }(\theta ^{\prime },\varphi ^{\prime
})Y_{lm}(\theta ,\varphi ).  \label{024}
\end{equation}
The parameters ${\bf (}k,\theta ^{\prime },\varphi ^{\prime })$ and ${\bf (}%
r,\theta ,\varphi )$ denote the corresponding components of ${\bf k}$ and $%
{\bf r}$ in spherical coordinates respectively. Using the same procedure as
in Eqs. (\ref{010})$\sim $(\ref{014}), we combine the nonlocal flux effect
into the partial wave expansion, and obtain the result 
\begin{equation}
e^{i{\bf k\cdot x}}e^{\frac{ie}{\hbar c}\int_{C}^{{\bf x}}{\bf A(\tilde{x}}%
)\cdot d{\bf \tilde{x}}}=\sum_{q=0}^{\infty }\sum_{m=-\infty }^{\infty
}(2\alpha +1)i^{\alpha }j_{\alpha }(kr ){\cal Y}_{qm}^{\ast }(\theta
^{\prime },\varphi ^{\prime }){\cal Y}_{qm}(\theta ,\varphi ).  \label{025}
\end{equation}
By employing approximations of spherical Bessel functions [see Eq. (\ref{041}%
)], 
\begin{equation}
j_{\alpha }(kr )\stackrel{r \rightarrow \infty }{\longrightarrow }%
\frac{1}{kr }\sin (kr -\alpha \pi /2),  \label{026}
\end{equation}
\begin{equation}
n_{\alpha }(kr )\stackrel{r \rightarrow \infty }{\longrightarrow }-%
\frac{1}{kr }\cos (\alpha \pi )\cos (kr +\alpha \pi /2),  \label{027}
\end{equation}
we can find that 
\begin{equation}
R_{\alpha k}(r )\stackrel{r \rightarrow \infty }{\longrightarrow }%
\frac{A_{\alpha }(k)}{kr}\left\{ \sin \left[ kr-\alpha \pi /2+\delta
_{\alpha }(k)\right] -\sin (\alpha \pi )\sin \delta _{\alpha }(k)\sin
(kr+\alpha \pi /2)\right\} .  \label{028}
\end{equation}
Substituting the result for $R_{\alpha k}(r )$ in (\ref{021}), and
comparing both asymptotic forms of (\ref{021}) and (\ref{023}), the
scattering amplitude is found to be 
\begin{equation}
f(\theta ,\varphi )=\frac{1}{k}\sum_{q=0}^{\infty }\sum_{m=-\infty }^{\infty
}\left( 2\alpha +1\right) \left[ \frac{e^{i\delta _{\alpha }}\sin \delta
_{\alpha }\cos ^{2}(\alpha \pi )}{1-e^{i(\delta _{\alpha }-\alpha \pi )}\sin
\delta _{\alpha }\sin (\alpha \pi )}\right] {\cal Y}_{qm}^{\ast }(\theta
^{\prime },\varphi ^{\prime }){\cal Y}_{qm}(\theta ,\varphi ).  \label{029}
\end{equation}
Here $(\theta ^{\prime },\varphi ^{\prime })$ is the incident direction of a
charged particle, and $(\theta ,\varphi )$ is the scattering direction. It
is easy to see that when the magnetic flux disappears, with $\theta ^{\prime
}=0$, i.e. ${\bf k}$ is along the z-axis, and $P_{l}(1)=1$, the result
reduces to the well-known amplitude 
\begin{equation}
f(\theta )=\frac{1}{k}\sum_{l=0}^{\infty }\left( 2l+1\right) e^{i\delta
_{l}}\sin \delta _{l}P_{l}(\cos \theta )  \label{029a}
\end{equation}
Let us consider the case of the incident direction perpendicular to the
magnetic flux, i.e. $(\theta ^{\prime }=\pi /2,\varphi ^{\prime }=0)$. We
have the function 
\begin{equation}
{\cal Y}_{qm}(\pi /2,0)=\sqrt{\frac{\Gamma (q+1)\Gamma (\alpha +\left| m+\mu
_{0}\right| +1)}{\Gamma ^{2}(\alpha +1)}}\left( \frac{1}{2}\right) ^{\left|
m+\mu _{0}\right| }P_{q}^{(\left| m+\mu _{0}\right| ,\left| m+\mu
_{0}\right| )}(0).  \label{030}
\end{equation}
With the help of the formulas (p.218$\sim $219 \cite{19}) 
\begin{equation}
P_{q}^{(\beta ,\beta )}(z)=\frac{\Gamma (2\beta +1)\Gamma (q+\beta +1)}{%
\Gamma (\beta +1)\Gamma (q+2\beta +1)}C_{q}^{\beta +1/2}(z),  \label{031}
\end{equation}
\begin{equation}
C_{q}^{\beta +1/2}(0)=\left\{ 
\begin{array}{l}
0\text{ \ if \ }q=\text{odd numbers} \\ 
(-1)^{\tilde{q}}\frac{\Gamma (\tilde{q}+\beta +1/2)}{\Gamma (\beta
+1/2)\Gamma (\tilde{q}+1)}\text{ if }q=\text{even numbers}
\end{array}
\right. ,  \label{031b}
\end{equation}
here $\tilde{q}\equiv q/2=0,1,2,\cdots $, and $C_{q}^{\beta +1/2}(z)$ is the
Gegenbauer polynomials, we can find that $P_{q}^{(\beta ,\beta )}(0)=0$ if $%
q=$odd numbers, and 
\[
P_{q}^{(\beta ,\beta )}(0) 
\]
\begin{equation}
=(-1)^{\tilde{q}}\frac{\Gamma (2\beta +1)\Gamma (2\tilde{q}+\beta +1)\Gamma (%
\tilde{q}+\beta +1/2)}{\Gamma (\beta +1)\Gamma (2\tilde{q}+2\beta +1)\Gamma
(\beta +1/2)\Gamma (\tilde{q}+1)},\text{ if\ }q=\text{even numbers,}
\label{032}
\end{equation}
where $\beta \equiv \left| m+\mu _{0}\right| $. Thus the function ${\cal Y}%
_{qm}(\pi /2,0)$ is given by 
\begin{equation}
{\cal Y}_{qm}(\pi /2,0)=(-1)^{\tilde{q}}\frac{1}{\sqrt{\pi }}\sqrt{\frac{%
\Gamma (\tilde{q}+1/2)\Gamma (\tilde{q}+\beta +1/2)}{\Gamma (\tilde{q}+\beta
+1)\Gamma (\tilde{q}+1)}}.  \label{033}
\end{equation}
In most cases, the total cross section of our major concern is defined by $%
\sigma _{t}=\int \sigma (\theta ,\varphi )d\Omega $, where $d\Omega $ is the
solid angle. By employing (\ref{016}), the partial wave representation of
total cross section for a charged particle scattered by a short range
potential plus the nonlocal AB effect is given by 
\begin{equation}
\sigma _{t}=\frac{4\pi }{k^{2}}\sum_{\tilde{q}=0}^{\infty }\sum_{m=-\infty
}^{\infty }F_{\tilde{q}m}(\delta _{\tilde{\alpha}})  \label{034}
\end{equation}
with 
\[
F_{\tilde{q}m}(\delta _{\tilde{\alpha}})=\left[ \frac{(2\tilde{\alpha}%
+1)\sin ^{2}\delta _{\tilde{\alpha}}\cos ^{4}(\tilde{\alpha}\pi ){\cal Y}_{%
\tilde{q}m}^{2}}{1-2\sin \delta _{\tilde{\alpha}}\sin (\tilde{\alpha}\pi
)\cos (\tilde{\alpha}\pi -\delta _{\tilde{\alpha}})+\sin ^{2}\delta _{\tilde{%
\alpha}}\sin ^{2}(\tilde{\alpha}\pi )}\right] , 
\]
where we have defined $\tilde{\alpha}\equiv (2\tilde{q}+\beta )$, and 
\begin{equation}
{\cal Y}_{\tilde{q}m}^{2}\equiv \frac{\Gamma \left( \tilde{q}+1/2\right)
\Gamma (\tilde{q}+\beta +1/2)}{\left[ \pi \Gamma \left( \tilde{q}+1\right)
\Gamma (\tilde{q}+\beta +1)\right] }.  \label{035}
\end{equation}
It is obvious that the cross section is completely determined by the
scattering phase shifts which are concluded by the potential of different
types. Furthermore, when a nonlocal AB magnetic flux exists, both the phase
shift and the cross section are affected globally. A relation between the
total cross section $\sigma _{t}$ and the scattering amplitude is obtained
if we set $\varphi =0$, and then take the imaginary part. It gives $\sigma
_{t}=\left( 4\pi /k\right) 
\mathop{\rm Im}%
f(\theta =\pi /2,\varphi =0)$. This is the optical theorem and is
essentially a consequence of the conservation of particles. For the
scattering of identical bosons (fermions) carrying the magnetic flux, the
differential cross section is given by $\sigma (\theta ,\varphi )=\left|
f(\theta ,\varphi )\pm f(\pi -\theta ,\varphi +\pi )\right| ^{2}$, where the
plus sign is for bosons as usual. The total cross sections are given by the
integral $\int_{-\pi }^{\pi }\sigma (\theta ,\varphi )d\Omega $, which yield 
\begin{equation}
\sigma _{t}({\rm bosons})=\frac{16\pi }{k^{2}}\sum_{\tilde{q}=0}^{\infty
}\sum_{m=-\infty ,{\rm even}}^{\infty }F_{\tilde{q}m}(\delta _{\tilde{\alpha}%
}),  \label{036}
\end{equation}
and 
\begin{equation}
\sigma _{t}({\rm fermions})=\frac{16\pi }{k^{2}}\sum_{\tilde{q}=0}^{\infty
}\sum_{m=-\infty ,{\rm odd}}^{\infty }F_{\tilde{q}m}(\delta _{\tilde{\alpha}%
}).  \label{037}
\end{equation}
Here the subscript ``odd'' (``even'') is used to indicate the summation over
odd (even) numbers only.

\section{Anomalous Cross Section induced by Quantum Interference}

As a realization of the nonlocal influence of the AB flux on the cross
section, let us consider a charged particle scattered by a hard sphere
potential and a magnetic flux. The potential is given by $V(r)=\infty ,$ for 
$r\leq a$, and $V(r)=0,$ for $r\leq a$. Using the boundary condition of the
wave function $R_{k\alpha }(a^{+})=0$, we find that the phase shift is given
by 
\begin{equation}
\tan \delta _{\tilde{\alpha}}=j_{\tilde{\alpha}}(ka)/n_{\tilde{\alpha}}(ka).
\label{038}
\end{equation}
Substituting this expression into (\ref{034}), the total cross section is
found to be 
\begin{equation}
\sigma _{t}=\frac{4\pi }{k^{2}}\sum_{\tilde{q}=0}^{\infty }\sum_{m=-\infty
}^{\infty }\frac{(2\tilde{\alpha}+1)\cos ^{2}(\tilde{\alpha}\pi )J_{\tilde{%
\alpha}+1/2}^{2}(ka){\cal Y}_{\tilde{q}m}^{2}}{J_{_{\tilde{\alpha}%
+1/2}}^{2}(ka)+J_{_{-\tilde{\alpha}-1/2}}^{2}(ka)+2\sin (\tilde{\alpha}\pi
)J_{\tilde{\alpha}+1/2}(ka)J_{_{-\tilde{\alpha}-1/2}}(ka)}.  \label{039}
\end{equation}
To obtain the result, we have applied the following relations between the
Bessel functions and spherical Bessel functions 
\begin{equation}
j_{\nu }(z)=\sqrt{\frac{\pi }{2z}}J_{\nu +1/2}(z),  \label{040}
\end{equation}
and 
\begin{equation}
n_{\nu }(z)=\left[ \cos (\nu +1)\pi \right] \sqrt{\frac{\pi }{2z}}J_{-\nu
-1/2}(z).  \label{041}
\end{equation}
The asymptotic behavior in (\ref{027}) can be found by the equality. Note
that the result will reduce to the pure hard sphere case 
\begin{equation}
\sigma _{t}=\frac{4\pi }{k^{2}}\sum_{l=0}^{\infty }\frac{(2l+1)j_{l}^{2}(ka)%
}{j_{l}^{2}(ka)+n_{l}^{2}(ka)}  \label{042}
\end{equation}
if the flux disappears, i.e. $\mu _{0}=0$. In this case the low energy limit 
$k\rightarrow 0$ (assuming the radius $a$ is finite) of phase shift can be
found by the asymptotic expansion of Bessel functions, which yields 
\begin{equation}
\tan \delta _{l}=j_{l}(ka)/n_{l}(ka)\stackrel{k\rightarrow 0}{%
\longrightarrow }-\frac{(ka)^{2l+1}}{[(2l-1)!!]^{2}(2l+1)}.  \label{043}
\end{equation}
Obviously, only index $l=0$ survives. The phase shift becomes 
\begin{equation}
\tan \delta _{0}(k)=j_{0}(ka)/n_{0}(ka)\approx -ka<0.  \label{022a}
\end{equation}
So the total cross section 
\begin{equation}
\sigma _{t}\approx \frac{4\pi }{k^{2}}\sin ^{2}\delta _{0}\approx \frac{4\pi 
}{k^{2}}\delta _{0}^{2}\approx 4\pi a^{2}.  \label{044}
\end{equation}
At the high energy limit $k\rightarrow \infty $, we may use the formulas of
spherical Bessel functions of the large argument to turn Eq. (\ref{042})
into 
\[
\sigma _{t}\approx \frac{4\pi }{k^{2}}\sum_{l=0}^{[ka]}(2l+1)\sin
^{2}(ka-l\pi /2) 
\]
\[
=\lim_{ka\rightarrow \infty }\frac{4\pi }{k^{2}}\left\{ \sum_{l=0,2,\cdots
}^{[ka]}(2l+1)\sin ^{2}\left( ka\right) +\sum_{l=1,3,\cdots
}^{[ka]}(2l+1)\cos ^{2}\left( ka\right) \right\} 
\]
\begin{equation}
=\lim_{ka\rightarrow \infty }\frac{4\pi }{k^{2}}\left\{ \frac{ka(ka+1)}{2}%
\sin ^{2}\left( ka\right) +\frac{(ka-1)(ka+2)}{2}\cos ^{2}\left( ka\right)
\right\} \approx 2\pi a^{2}.  \label{045}
\end{equation}
The numerical result for $\alpha $ with noninteger value is plotted in Fig.
1, where the normalization $\sigma _{0}$ is chosen as $2\pi a^{2}$. There
are two main results which are caused by the quantum interference of the AB
effect: (1) The cross section $\sigma _{t}$ is drastically suppressed at the
low energy limit (equivalently, the short range potential), say $ka\leq 1$,
at quantized magnetic flux $\Phi =(2n+1)\Phi _{0}/2$, $n=0,1,2,\cdots $,
with $\Phi _{0}$ periodicity as shown in Fig. 1 and Fig. 2. (2) A more
interesting consideration is given by the scattering of identical particles
simulated by the hard spheres carrying the magnetic flux. In Fig. 3, we plot
the total cross sections of identical bosons carrying the magnetic flux via
Eq. (\ref{036}). The outcome shows that the cross section approaches zero ($%
\sigma _{t}\rightarrow 0$) when the value $ka\rightarrow 0$ if the magnetic
flux is at quantized value $(2n+1)\Phi _{0}$. On the contrary, if the
magnetic flux is equal to $2n\Phi _{0}$, the cross section becomes maximum
and the effect of magnetic flux disappears. Since the decay rate of a
current ${\bf j}$ traveling a distance ${\bf x}$ is given by ${\bf j(x)=j(}0%
{\bf )}\exp (-\sigma _{t}n_{0}{\bf x})$, where $n_{0}$ is the number of the
scattering center, the total cross section $\sigma _{t}\rightarrow 0$ at the
low energy limit at $\Phi =(2n+1)\Phi _{0}$ means that the resistance $%
R\rightarrow 0$ and results in the persistence of current. This phenomenon
is consistent with the picture of composite boson in fractional quantum Hall
states located at the filling factor with odd denominator such as $\nu =1/3$%
. The composite boson is pictured by an electron carrying the quantized
magnetic flux $\Phi =(2n+1)\Phi _{0}$. It dictates the quantized Hall states
which exhibit the perfect conduction in the longitudinal direction, i.e. the
resistance originated from the collisions between composite bosons disappear 
\cite{3}. The global structure of the total cross section is given by $2\Phi
_{0}$ periodicity as shown in Fig. 4. In the case of identical fermions, the
total cross section $\sigma _{t}\rightarrow 0$ is found at the quantized
magnetic flux $\Phi =2n\Phi _{0}$ as shown in Fig. 5. Such effect is
consistent with the model of composite fermion in the quantum Hall state
located at the filling factor with even denominator $\nu =5/2$. The
composite fermion is described by an electron carrying the quantized
magnetic flux $\Phi =2n\Phi _{0}$. In Ref. \cite{21}, a quantitative
explanation of quantum Hall state at the filling factor $\nu =5/2$ is given
by the existence of a shorter range potential between the composite fermions
than the case of the filling factor $\nu =1/2$. Here we can see that, in
Fig. 5, a sufficiently short range potential, say $ka<0.5$, between the
fermions carrying the quantized magnetic flux $\Phi =2n\Phi _{0}$ will cause
negligible cross section and thus agree with the composite fermions model.
Similar to the boson case, the oscillating period is given by $2\Phi _{0}$
as shown in Fig. 6.

\section{Discussion}

1. {\it Symmetries}

When the incident direction is perpendicular to the magnetic flux,i.e. $%
(\theta ^{\prime }=\pi /2,\varphi ^{\prime }=0)$, we see from (\ref{031b})
that $q$ must be equal to even numbers so that these channels have
nonvanishing contributions. In this case we have 
\begin{equation}
P_{2\tilde{q}}^{(\beta ,\beta )}(\cos (\pi -\theta ))=P_{2\tilde{q}}^{(\beta
,\beta )}(\cos \theta ).  \label{b1}
\end{equation}
On the other hand, from (\ref{022}) we have 
\begin{equation}
{\cal Y}_{qm}(\pi -\theta ,\varphi )={\cal Y}_{qm}(\theta ,\varphi ).
\label{b2}
\end{equation}
These two equalities give us the relation in (\ref{029}) 
\begin{equation}
f(\theta ,\varphi )=f(\pi -\theta ,\varphi )  \label{b3}
\end{equation}
which means that the amplitude, and thus the cross section, is symmetric
about the $x$-$y$ plane. If we make the condition $\varphi \rightarrow
-\varphi ,$ which is equivalent to $m\rightarrow -m$, the effect is equal to
reverse the direction of flux form $+z$ to $-z$ since $\left| m+\mu
_{0}\right| \rightarrow \left| -m+\mu _{0}\right| =\left| m-\mu _{0}\right| $%
.

2. {\it Anomalous Cross Section Induced by the Quantum Interference}

By way of Fig. 1, we see that the total cross section may be anomalous due
to the quantum interference which provides another possibility to explain
the depression of the total cross section discussed in recent papers \cite{9}%
, where the quantum entanglement is supposedly responsible for the
suppression of total cross section in the condensed matter. However, issues
do exist regarding that the lifetime of the entanglement in condensed system
is much shorter than the present-day time-resolution techniques can resolve,
and therefore it is commonly expected to have no experimental significance.

3. {\it The effects of magnetic flux and dimensions}

The quantum interference features in Fig. 1$\sim $Fig. 6 were observed in 
\cite{23} where a two dimensional partial wave analysis of scattering with
nonlocal AB effect was constructed, and a ``hard disk'' with the AB magnetic
flux was used to simulate the dynamics of a charged particle with magnetic
flux. Although the dominate picture of the quantum interference can be found
in the hard disk model, it is somewhat too simple to yield a ``wave packet
like'' object. In the paper, with the hard sphere model, we see from Fig. 1$%
\sim $Fig. 6 that quantum interference features at the quantized magnetic
flux $\Phi =(2n+1)\Phi _{0}/2$, $2n\Phi _{0}$, and $(2n+1)\Phi _{0}$ are
apparent.

4. {\it Extension the potential to more general case}

Although in the procedure of our proof we assume $V(r)=0$ for $r>a$, we do
not specify the radius $a$ beyond which $V(r)=0$. Hence we expect that the
theorem given in the article should be valid for a very general potential as
long as the potential decrease rapidly enough when $r\rightarrow \infty $.

5. {\it A possible experimental test}

In Ref. \cite{22}, a general fractional (non-quantized) magnetic flux is
observed in the superconducting film. Because of the inevitable pinning of
flux in superconductor, the flux finally attaches to the defect or impurity
such that they become models of a finite range interaction with flux as
mentioned in point 4. The system scattered by the other low energy charged
particle can be as the test ground of the anomalous cross section presented
in the paper.
\centerline{ACKNOWLEDGMENTS} 
\center{The author would like to thank
professor Pi-Guan Luan for helpful discussions, Professors Der-San Chuu, and Jang-Yu Hsu for reading the manuscript.}
\newpage
\begin{figure}[hbt]\includegraphics[width=2.8in]{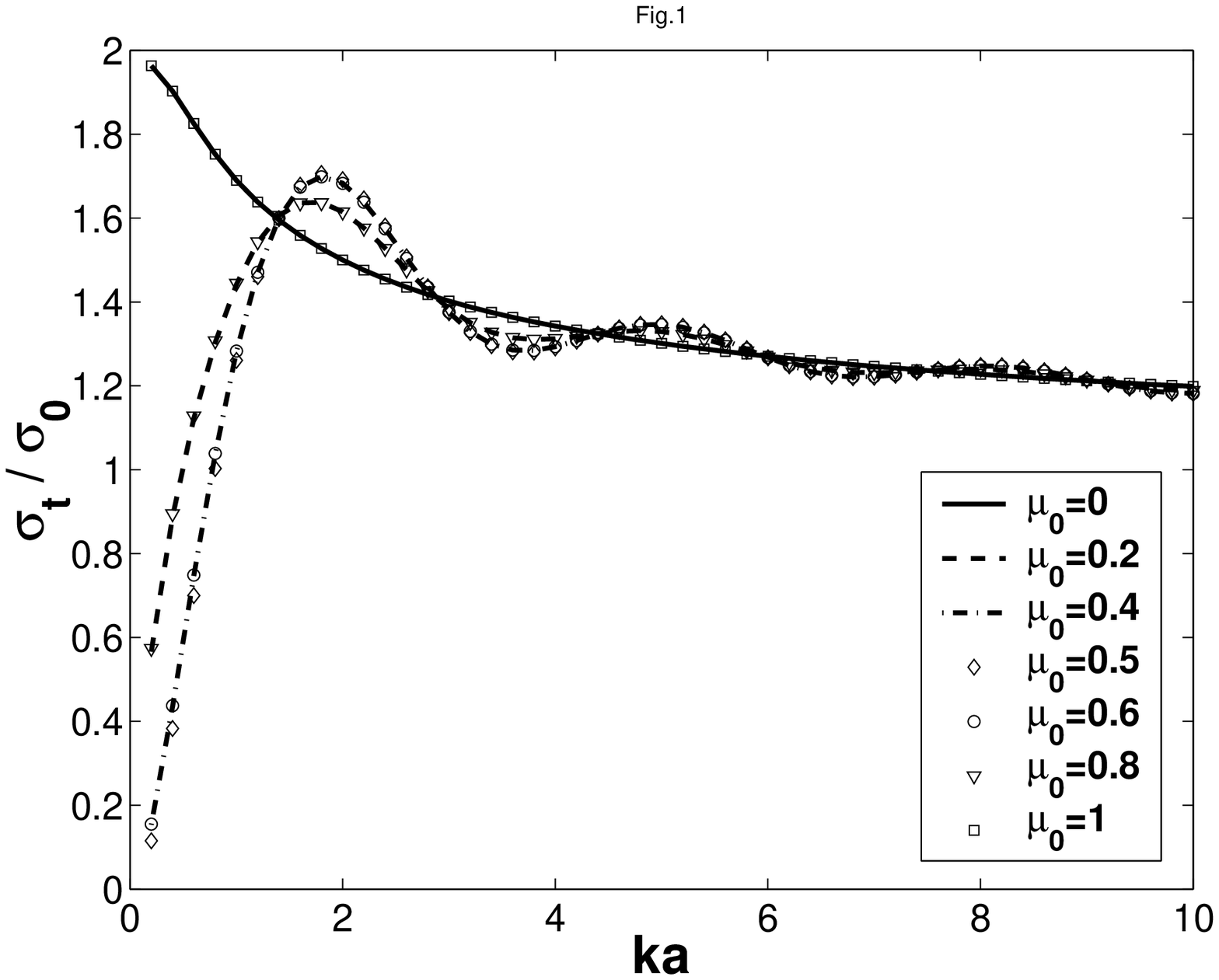}
\caption{ The total cross section for a charged particle scattered by a hard
sphere with radius $a$ and a magnetic flux along the z-axis. The
normalization $\protect\sigma_0=2 \pi a^2$ has been selected. Due to the existence
of magnetic flux, at the limit of the long wave
(equivalently, the short range potential), say $ka \leq 1$, the total corss section is drastic suppressed
at quantized magnetic flux $\Phi =(2n+1)\Phi _{0}/2$, where $n=0,1,2,\cdots $, with $\Phi _{0}$ periodicity, see Fig. 2.
The magnetic flux effect disappears when
the flux is quantized at $\Phi=n\Phi _{0}$.}
\end{figure}

\begin{figure}[hbt]\includegraphics[width=2.8in]{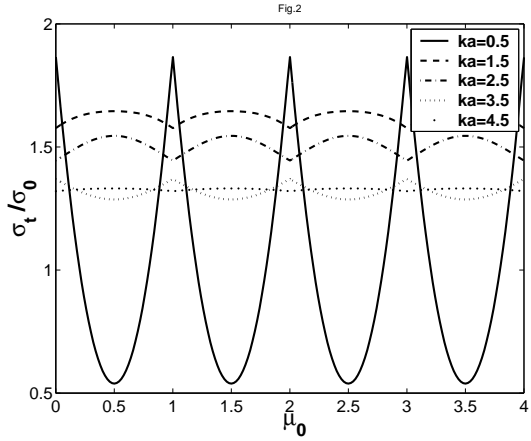}
\caption{Periodic structures of total cross sections of
a charged particle scattered by
a hard sphere plus a magnetic flux along the z-axis.
At quantized values of magnetic flux $\Phi=(2n+1)\Phi _{0}/2$, $n=0,1,2,\cdots $, the
cross section reduces to the minimum for $ka \leq 0.5$.}
\end{figure}

\begin{figure}[hbt]\includegraphics[width=2.8in]{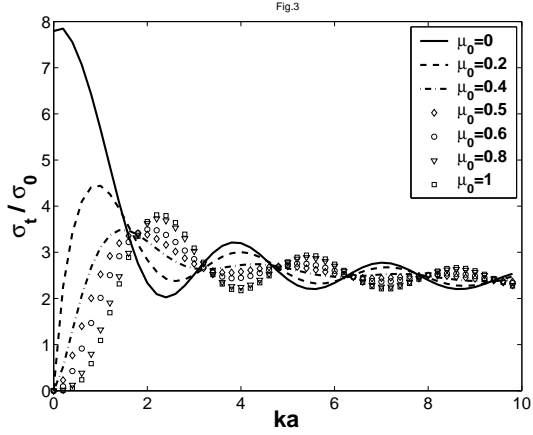}
\caption{Total cross sections for identical bosons carrying the
magnetic flux with various $\protect\mu_0$. The cross section at the
long wave length limit (equivalently, the sufficient short range potential), say $ka \leq 0.5$,
approaches zero at the quantized magnetic flux
$\Phi=(2n+1)\Phi _{0}$.
On the contrary, the cross section becomes maximum and the effect of magnetic flux disappears
when $\Phi=2n\Phi _{0}$.
The periodic structure is $2\Phi _{0}$ as shown in Fig. 4.}
\end{figure}

\begin{figure}[hbt]\includegraphics[width=2.8in]{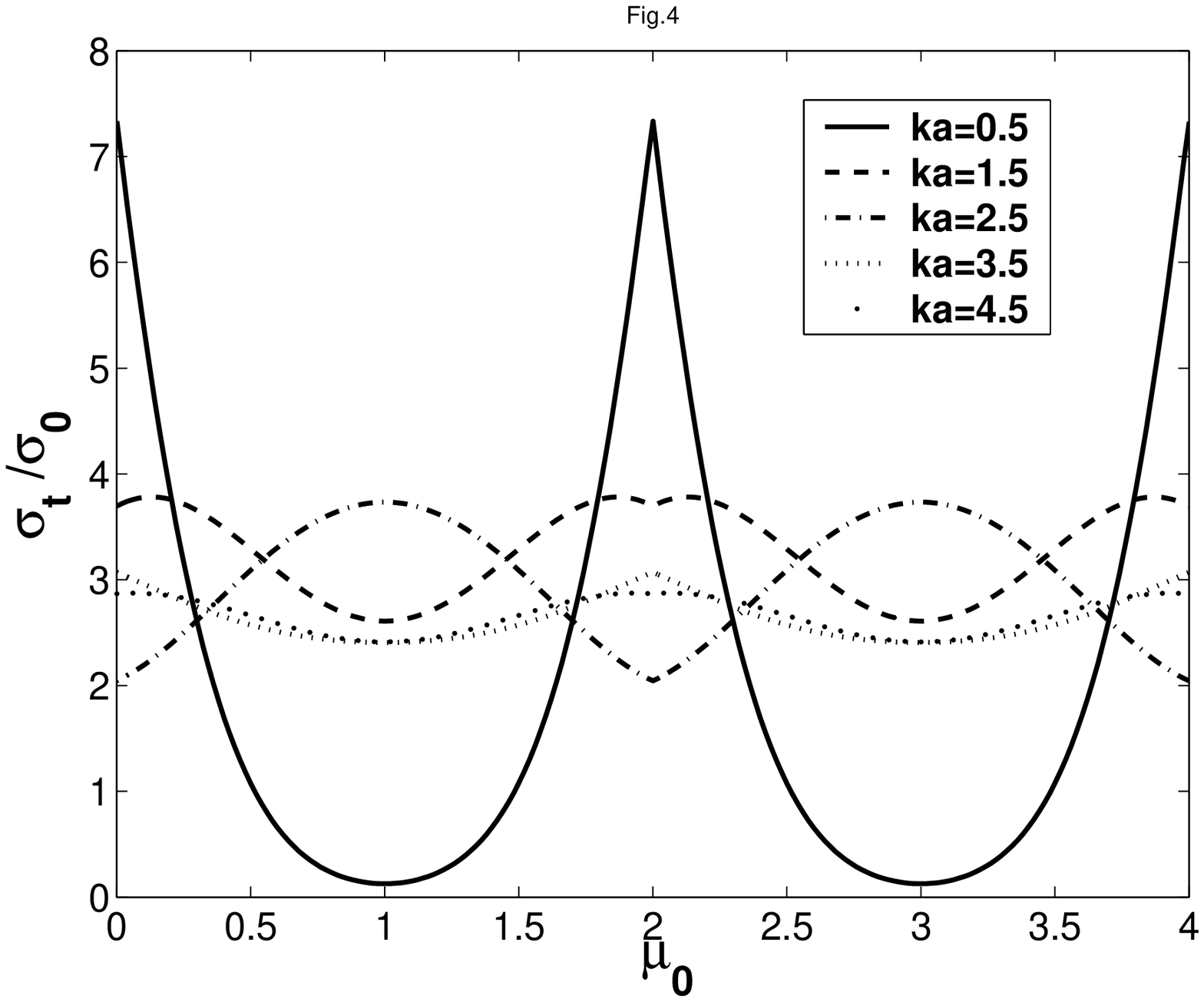}
\caption{Periodic structures of cross sections of
identical bosons carrying the magnetic flux. The cross section approaches zero
when the magnetic flux is quantized at $\Phi=(2n+1)\Phi_{0}$ for $ka \leq 0.5$.}
\end{figure}

\begin{figure}[hbt]\includegraphics[width=2.8in]{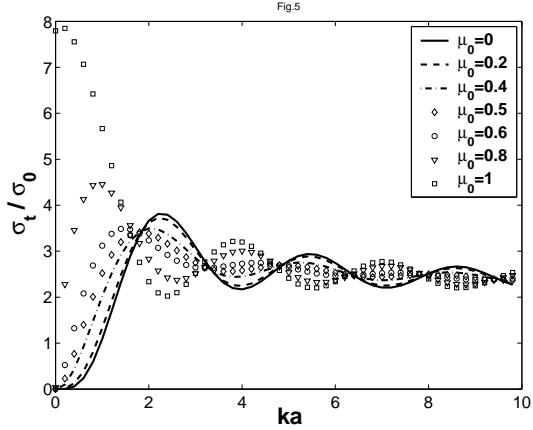}
\caption{Total cross sections of identical fermions carrying the
magnetic flux with various $\protect\mu_0$. The
cross section approaches zero for $ka \leq 0.5$ when the flux becomes
$2n\Phi _{0}$. The magnetic flux effect disappears
when the magnitude of flux is at $(2n+1)\Phi _{0}$.
The global periodic structures in cross sections is $2\Phi _{0}$ as shown in Fig. 6.}
\end{figure}

\begin{figure}[hbt]\includegraphics[width=2.8in]{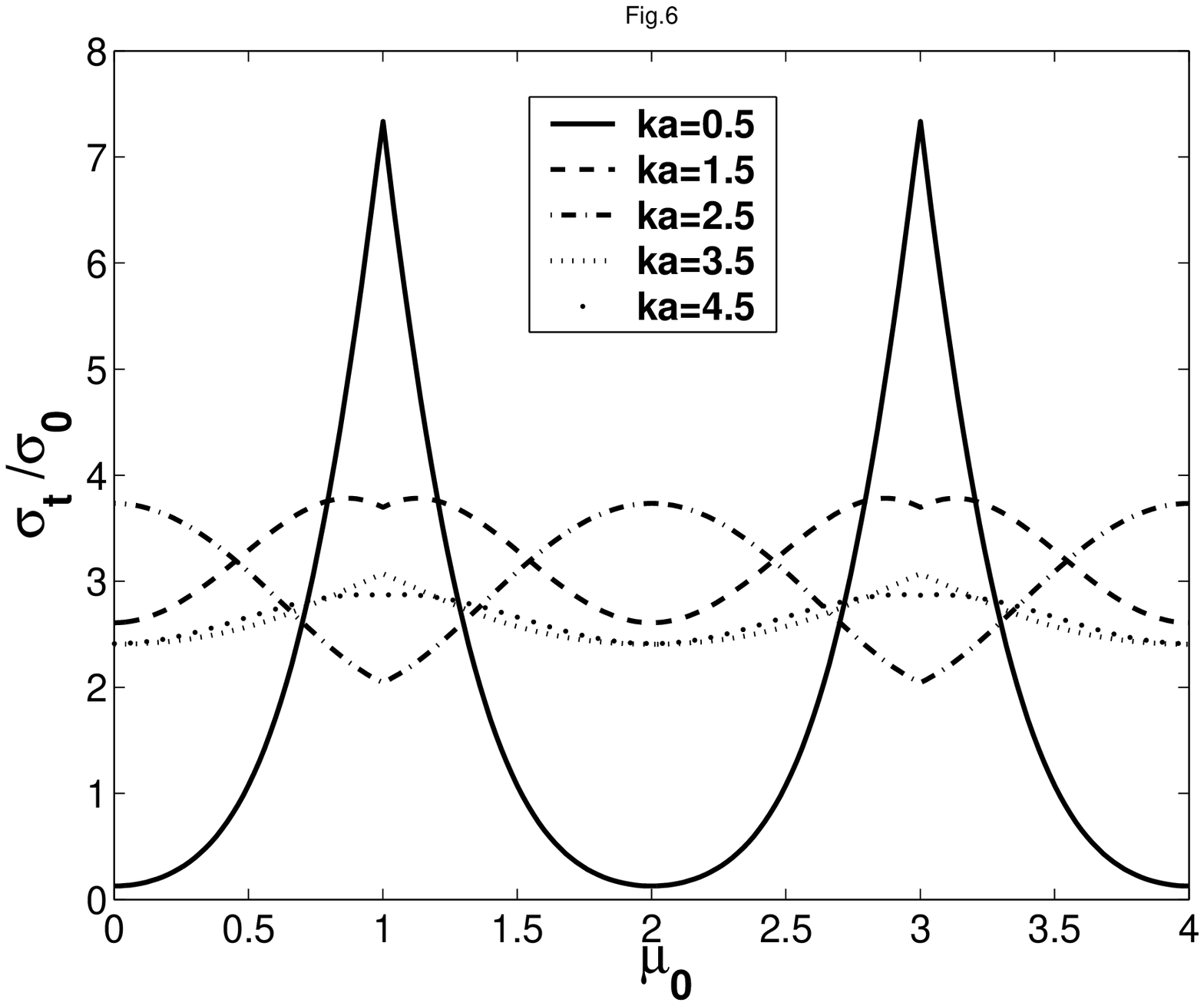}
\caption{Periodic structures of total cross sections
for identical fermions carrying the magnetic flux. The cross section approaches
zero when the magnetic flux is quantized at $\Phi=2n\Phi _{0}$ for $ka \leq 0.5$.}
\end{figure}
\end{document}